\documentclass[aps,prl,preprint,groupedaddress,nofootinbib]{revtex4-1}

\usepackage{graphicx}
\begin{document}


\title{\Large Freeze-in  Dark Matter from a Minimal B-L Model  and Possible Grand Unification}


\author{\bf Rabindra N. Mohapatra$^a$}
\author{Nobuchika Okada$^b$}
\affiliation{}
\affiliation{$^a$ Maryland Center for Fundamental Physics and Department of Physics, University of Maryland, College Park, Maryland 20742, USA}
\affiliation{$^b$ Department of Physics and Astronomy, University of Alabama, Tuscaloosa, Alabama 35487, USA}



\begin{abstract} 
We show that a minimal local $B-L$ symmetry extension of the standard model can provide a unified description of both neutrino mass and dark matter.  In our model,  $B-L$ breaking is responsible for neutrino masses via the seesaw mechanism, whereas the real part of the $B-L$ breaking Higgs field (called $\sigma$ here)  plays the role of a freeze-in dark matter candidate for a wide parameter range. Since the $\sigma$-particle is unstable, for it to qualify as dark matter, its lifetime must be longer than $10^{25}$ seconds implying that the $B-L$ gauge coupling must be very small. This in turn implies that  the dark matter relic density must arise from the freeze-in mechanism.    The dark matter lifetime bound combined with dark matter relic density gives a lower bound on the $B-L$ gauge boson mass in terms of the dark matter mass. We point out parameter domains where the dark matter mass can be both in the keV to MeV range as well as in the PeV range. We discuss ways to test some parameter ranges of this scenario in collider experiments. Finally, we show that if instead of $B-L$, we consider the extra $U(1)$ generator to be $-4I_{3R}+3(B-L)$,  the basic phenomenology remains unaltered and for certain gauge coupling ranges, the model can be embedded into a five dimensional $SO(10)$ grand unified theory.


\end{abstract}

\maketitle

\section{1. Introduction}
If small neutrino masses arise via the seesaw mechanism \cite{seesaw1,seesaw2,seesaw3,seesaw4,seesaw5}, 
the addition of a local $B-L$ symmetry \cite{marshak1,marshak2} to the standard model (SM) provides a minimal scenario for beyond the standard model (BSM) physics to achieve this goal. 
There are  two possible classes of $B-L$ models: one where the $B-L$ generator contributes to the electric charge~\cite{marshak1,marshak2,davidson} and another where it does not~\cite{BL1, BL2, BL3}. In the first case, the $B-L$ gauge coupling $g_{BL}$ has a lower limit whereas in the second case it does not and therefore can be arbitrarily small.  There are constraints on the allowed ranges of $g_{BL}$ from different observations~\cite{heeck,bauer} in the second case depending on whether there is or is not a dark matter particle in the theory.

In Refs.~\cite{nobu, heeba}, it was shown that if we added a $B-L$ charge carrying vector-like fermion to the minimal $B-L$ model and want it to play the role of dark matter, new constraints emerge. In this note, we discuss an alternative possibility with the following new results. First is that the minimal version of the $B-L$ model itself, without any extra particles, can  provide a dark matter (DM) candidate. The DM turns out to be the real part (denoted here as $\sigma$) of the complex $B-L=2$ Higgs field, that breaks $B-L$ and gives mass to the right handed neutrinos in the seesaw formula. Even though this particle is not stable,  there are certain allowed parameter ranges of the model, where its lifetime can be so long  that it can play the role of a decaying dark matter. We isolate this parameter range and
 show that in this case,  the freeze-in mechanism~\cite{hall} can generate its relic density. We find  this possibility to be interesting since it unifies both neutrino masses and dark matter in a single minimal framework. 
We show how a portion of the parameter range of the model suggested by the dark matter possibility, can be probed by the recently approved FASER experiment at the LHC \cite{faser} and other Lifetime Frontier experiments. 

We then show that if we replace the $B-L$ symmetry by $\tilde{I}\equiv -4I_{3R}+3(B-L)$ (where $I_{3R}$ is the right handed weak isospin), the dark matter phenomenology remains largely unchanged and the model can be embedded into the $SO(10)$ grand unified theory in five space-time dimensions. Such a symmetry breaking of $SO(10)$ to $SU(5)\times U(1)_{\tilde{I}}$ has already been shown to arise from a symmetry breaking 
by a particular alignment for a vacuum expectation value (VEV) of a {\bf 45}-dimensional Higgs field \cite{diluzio}.

This paper is organized as follows: in Sec.~2 after briefly introducing the model, we discuss the lifetime of the $\sigma$ dark matter and its implications. In  Sec.~3, we discuss the small gauge coupling $g_{BL}$ range where the dark matter lifetime is long enough for it to play the role of dark matter. 
In Sec.~4, we show how freeze in mechanism determines the relic density of dark matter and its implications for the allowed parameter range of the model. 
We also discuss how to test this model at the FASER and other Lifetime Frontier experiments. 
In Sec.~5, we show that this model can also accommodate a PeV dark matter. 
In Sec.~6,  we discuss the $SO(10)$ embedding of the closely allied model and in Sec.~7, 
we conclude with some comments and other implications of the model.

\section{2. Brief overview of the model} 
Our model is based on the $U(1)_{B-L}$ extension of the SM with gauge quantum numbers under $U(1)_{B-L}$ determined by the baryon or lepton number of the particles. The gauge group of the model is $SU(3)_c \times SU(2)_L\times U(1)_Y\times U(1)_{B-L}$, where $Y$ is the SM hypercharge. We need three right handed neutrinos (RHNs) with $B-L=-1$ to cancel the $B-L$ anomaly. The RHNs being SM singlets do not contribute to SM anomalies. The electric charge formula in this case is same as in the SM i.e. $Q=I_{3L}+\frac{Y}{2}$. 

We break $B-L$ symmetry by giving a VEV to a $B-L=2$ SM neutral complex Higgs field $\Delta$ i.e. $\langle \Delta\rangle =v_{BL}/\sqrt{2}$. 
This gives Majorana masses to the right handed neutrinos ($N$) via the coupling $f N N \Delta$. 
The real part of $\Delta$ (denoted by $\sigma$)  is a physical field. Our goal in this paper is to show that $\sigma$ has the right properties to play the role of a dark matter of the universe. There are three challenges to achieving this goal: 

\noindent 
(i) The $\sigma$ field has couplings to the RHNs which in turn couples to SM particles providing a way for $\sigma$ to decay. 
Also, the $\sigma$ field has couplings to two $B-L$ gauge bosons ($Z_{BL}$) which in turn couple to SM fields providing another channel for $\sigma$ to decay. In the next section, we show that there are parameter regions of the model where these decay modes give a long enough lifetimes for $\sigma$, so that it can be a viable unstable dark matter in the universe.

\noindent 
(ii) The second challenge is that for $\sigma$ to be a sole dark matter, it must account for the total observed relic density of the universe $\Omega_{DM}h^2\simeq 0.12$ \cite{Planck2018}. We show in Sec.~4 that in the same parameter range, that gives rise to the long lifetime of $\sigma$, can also explain the observed relic density of dark matter via the freeze-in mechanism.

\noindent
(iii) The $\sigma$ field could mix with the standard model Higgs field $h$ via the potential term $\lambda^\prime H^\dagger H \Delta^\dagger \Delta$ after symmetry breaking. However, it turns out that if we set $\lambda^\prime =0$ at the tree level, it can be induced at the one-loop level by fermion  contributions and at the two-loop level from the top loop as shown in Ref.~\cite{BL3}. 
These induced couplings can be so small that they still lead to very long lifetimes for $\sigma$ in the parameter range of interest to us.

\section{ 3. Dark matter lifetime} 
As noted earlier in Sec.~2, the $\sigma$ field has couplings which could make it unstable and thereby disqualify it from being a dark matter. However, we will show that there is a viable parameter range of the model where this decay lifetime is longer than $10^{25}$ sec.~\cite{farinaldo} so that it can be a dark matter candidate. We discuss these two modes now:

\noindent (i) Decay mode {\bf $\sigma \to NN \to \ell f\bar{f}\ell f\bar{f}$}: the decay width for this process is estimated as 
\begin{eqnarray}
\Gamma_{NN}\simeq \frac{(f \, h_\nu^2 \, h^2_{SM})^2}{(4 \pi)^8} \frac{m^{13}_\sigma}{M^4_N \, m_h^8},
\end{eqnarray}
where $h_\nu$ is a neutrino Dirac Yukawa coupling, $h_{SM}$ is a Yukawa coupling of an SM fermion $f$, and $m_h=125$ GeV 
  is the SM Higgs boson mass. 
For a GeV mass $\sigma$ and TeV mass RHN, the lifetime of $\sigma $ turns out to be 
  $\tau_\sigma[{\rm sec}] \sim 10^{37}/(f^2 \, h_{SM}^4)$, which is quite consistent with the requirement for it to be a dark matter.
Here, we have used the seesaw formula $h_\nu^2 v_{EW}^2/M_N \simeq m_\nu$ with $v_{EW}=246$ GeV
  and a typical neutrino mass scale $m_\nu \simeq 0.1$ eV.

\noindent(ii) Decay mode {\bf $\sigma\to Z_{BL}Z_{BL}\to f\bar{f}f\bar{f}$}: the decay width for this process is
\begin{eqnarray}
\Gamma_{Z_{BL}Z_{BL}}\simeq \frac{(2 g_{BL})^4 \, v_{BL}^2 \,  g_{BL}^4 \, m_{\sigma}^7}{(4 \pi)^5  M^8_{Z_{BL}}}
=\frac{g_{BL}^6}{ 256 \pi^5}  \frac{m_\sigma^7}{M^6_{Z_{BL}}}. 
\end{eqnarray}
This mode is sensitive to the values of $g_{BL}$ as well as $M_{Z_{BL}}$. 
The estimate of $\tau_\sigma$ due to this decay mode is given by
\begin{eqnarray}
\tau_\sigma\simeq  5.2 \times 10^{-20} \left(\frac{1}{g_{BL}}\right)^6\left(\frac{{\rm 1 \, GeV}}{m_\sigma}\right)^7 \left(\frac{M_{Z_{BL}}}{{\rm 1\, GeV}}\right)^6~~{\rm sec.}
\label{LF}
\end{eqnarray}
Imposing $\tau_\sigma > 10^{25}$ sec., this puts an upper bound on the $g_{BL}$ as a function of $M_{Z_{BL}}$ and $m_\sigma$:
\begin{eqnarray}\label{lifetime}
g_{BL} \leq  4.2 \times 10^{-8} \left(\frac{M_{Z_{BL}}}{{\rm 1 \, GeV}}\right) \left(\frac{{\rm 1 \,GeV}}{m_\sigma}\right)^{7/6}
\label{LF2}
\end{eqnarray}
We find that the allowed regions where the $\sigma$ field can be a dark matter correspond to a very small $g_{BL}$ coupling. 
For instance, for $m_\sigma  \sim 1$ GeV and $M_{Z_{BL}} \sim 1$ TeV, we find that $g_{BL} \lesssim 4 \times 10^{-5}$. 

\noindent(iii) We now comment on the $\sigma$-Higgs mixing effect on the DM lifetime. 
To keep the lifetime  above limit $\tau_\sigma > 10^{25}$ sec., we set the tree-level  $H$-$\Delta$ coupling in the Higgs potential to zero so that $\sigma$ and the SM Higgs field $h$ do not mix at the tree level. This will, for example be true if the model becomes supersymmetric at a high scale. The $\sigma$-Higgs mixing in this case is loop induced as shown in Ref.~\cite{BL3} and for the parameter range of interest to us, can be small enough to satisfy the DM lifetime constraint as we show below. 

 For the case when $m_\sigma \leq m_h$, the dominant contribution to the loop induced mixing comes from a RHN fermion box diagram and the mixing angle can be estimated to be 
$\theta\sim \frac{f^2 h^2_\nu}{16\pi^2}\frac{v_{EW} v_{BL}}{m^2_h} 
\sim \frac{1}{16\pi^2}  \frac{m_\nu M_N^3}{v_{EW} m_h^2} \frac{2 g_{BL}}{M_{Z_{BL}}} $. 
Through this mixing, the DM particle can decay to a pair of SM fermions with a partial decay width of 
$\Gamma_{\sigma \to f \bar{f}} \sim \frac{\theta^2}{4 \pi} \left(\frac{m_f}{v_{EW}} \right)^2 m_\sigma$. 
The lifetime constraint then translates to a limit on $g_{BL}$ as follows:
\begin{eqnarray}
g_{BL} < 2.8 \times 10^{-6} \left(\frac{v_{EW}}{m_f}  \right) 
\left( \frac{1 \,{\rm GeV}}{m_\sigma} \right)^{1/2}
\left(\frac{1 \,{\rm  GeV}}{M_N}\right)^{3} \left(\frac{M_{Z_{BL}}}{1 \, {\rm GeV}}\right). 
\end{eqnarray}
With a suitable choice of $M_N (> m_\sigma)$, we can see that this limit is quite compatible
with our results shown in the right panel of Figs.~\ref{Fig:1}, Fig.~\ref{Fig:2} . 

For the case when $m_\sigma > m_h$, on the other hand, the DM particle can decay to a pair of Higgs doublets
through the mixing, and we  find that the loop induced mixing is not small enough to be consistent with the results 
shown in the right panels of Figs.~\ref{Fig:1} and \ref{Fig:3}.
In this case, we consider a cancellation of the mixing between the tree and loop levels contriburions.  

We will now explore whether for such small parametric values for $g_{BL}$, we can generate the observed dark matter relic density of the universe.

\section{4. Relic density} 
\subsection{4.1 Allowed range of $g_{BL}$ from pre-conditions to freeze-in}
First point to notice is that for GeV scale DM ($\sigma$), for values of $g_{BL}$ that satisfy the lifetime constraint, the $\sigma$ field is out of equilibrium from the SM particles. Therefore, the standard thermal freeze-out mechanism for creation of DM relic density does not apply and one has to explore the freeze-in mechanism. For this to work, we need the $Z_{BL}$ field, whose annihilation will produce the DM, to be in equilibrium with the SM fields. 
This question was explored in Ref.~\cite{nobu} and it was pointed out that the most efficient process for $Z_{BL}$ to be in equilibrium with SM particles is via the process $f\bar{f}\to Z_{BL}+\gamma$. The condition on $g_{BL}$ for this to happen  is  $g_{BL} > 2.7\times 10^{-8}\left(\frac{M_{Z_{BL}}}{{\rm 1\,  GeV}}\right)^{1/2}$.

An upper bound on $g_{BL}$ comes from the fact that the DM particle $\sigma$ is out of equilibrium in the early universe.
The first process to consider is $Z_{BL} Z_{BL} \leftrightarrow \sigma \sigma$ for which
  the out-of-equilibrium condition is given by $n_\sigma \langle \sigma v \rangle < H$,. 
  Here $n_\sigma \sim T^3$ is the number density of the DM $\sigma$, 
  $\langle \sigma v \rangle \sim g_{BL}^4/(4 \pi T^2)$, 
  and the Hubble parameter $H =\sqrt{\frac{\pi^2}{90} g_*} T^2/M_P$ with the reduced Planck mass $M_P=2.43 \times 10^{18}$ GeV 
  and the effective total number of relativistic degrees of freedom $g_*$ 
  (we set $g_*=106.75$ for the SM particle plasma in our analysis throughout this paper).     
Requiring that this inequality is satisfied until $T \sim M_{Z_{BL}}$, 
    we find that $g_{BL} < 6.4 \times 10^{-5}  \left(\frac{M_{Z_{BL}}}{\rm 1 \, GeV}\right)^{1/4}$. 
Combining with the equilibrium condition for $Z_{BL}$, we find that we have  to work in the range of $g_{BL}$ values
\begin{eqnarray}
    2.7\times 10^{-8}\left(\frac{M_{Z_{BL}}}{{\rm 1\,  GeV}}\right)^{1/2} < g_{BL} <  6.4 \times 10^{-5}  \left(\frac{M_{Z_{BL}}}{\rm 1 \, GeV}\right)^{1/4}. 
\label{TH}
\end{eqnarray}
to generate the relic density.

There is another upper bound on $g_{BL}$ that arises from the fact that the process $NN\to \sigma\sigma$ should also out of equilibrium. 
The reason is that in the early universe, the right handed neutrinos are always  in equilibrium with SM particles via processes 
such as $N+t \leftrightarrow  \nu+t$ etc.~and $N \leftrightarrow H \ell$ for $M_N > m_h$. If $NN \leftrightarrow \sigma\sigma$ is also in equilibrium, 
the freeze-in mechanism for relic density generation of $\sigma$ will not work. 
To get this upper bound on $g_{BL}$ using this condition, we use $n_\sigma \langle \sigma_{NN\to \sigma\sigma}v \rangle < H$ at $T \sim M_N$ and find
\begin{eqnarray}
\frac{1}{4\pi} \left(\frac{M^5_N}{v^4_{BL}} \right) <  \sqrt{\frac{\pi^2 }{90} g_*}\frac{M_N^2}{M_P}
\end{eqnarray}
Using $M_{Z_{BL}}=2g_{BL}v_{BL}$, this leads to
\begin{eqnarray}
%
g_{BL} < 3.2 \times 10^{-5}  \left(\frac{M_{Z_{BL}}}{\rm 1 \, GeV}\right)^{1/4} \left(\frac{M_{Z_{BL}}}{M_N}\right)^{3/4}. 
\label{THN}
\end{eqnarray}
Note that for $M_N \sim M_{Z_{BL}}$, this upper limit is about the same level as in Eq.~(\ref{TH}) 
so that indeed  the freeze-in mechanism is called for in creating the relic density build-up. 
In the following, we consider $M_N < M_{Z_{BL}}$, for which the upper bound is determined by the $B-L$ gauge interaction. 
Incidentally, we note that If $M_N < m_h$, the interactions of the RHNs with the SM particles are too week for them to be in thermal equilibrium, 
and the above discussion is not applicable.\footnote{
Note  also that, as a general possibility, if $M_N$ is greater than the reheating temperature after inflation ($T_{RH}$), 
the RHN is irrelevant to our DM physics discussion. 
}

\subsection{4.2 Relic density build-up}
In order to calculate the relic density build-up via the freeze-in mechanism, we solve the following Boltzmann equation
(defining $x=\frac{m_\sigma}{T}$):
\begin{eqnarray}
\frac{dY}{dx}\simeq \frac{ \langle \sigma v \rangle}{x^2}\frac{s(m_\sigma)}{H(m_\sigma)} Y^2_{eq}, 
\label{Boltzmann}
\end{eqnarray}
where $Y$ is the yield of the DM $\sigma$, $Y_{eq}$ is $Y$ if the DM $\sigma$ is in thermal equilibrium,  
  and $s(m_\sigma)$ and $H(m_\sigma)$ are the entropy density and the Hubble parameter, respectively,  
  evaluated at $T=m_\sigma$. 
For the DM particle creation process $Z_{BL} Z_{B} \to \sigma \sigma$, 
  we approximate $\langle \sigma v \rangle \simeq \frac{g_{BL}^4}{4 \pi T^2}= \frac{g^4_{BL}}{4 \pi} \frac{x^2}{m^2_\sigma}$. 
Note that this formula is applicable for $T\geq M_{Z_{BL}}\gg m_\sigma$. 
The reason for this is that for $T \leq M_{Z_{BL}}$, the number density of $Z_{BL}$ is Boltzmann suppressed
  and $\sigma$ particle creation stops. 
Using $\frac{S(m_\sigma)}{H(m_\sigma)}\simeq 14 \, m_\sigma M_P$ and $Y_{eq}\simeq 2.2 \times 10^{-3}$ 
  and integrating the above equation from $x_{RH}$ to $x$ (where $x_{RH}= m_\sigma/T_{RH}$ 
  with the reheating temperature after inflation $T_{RH} \gg M_{Z_{BL}}$), we obtain
\begin{eqnarray}
Y(x)-Y(x_{RH})\simeq 5.1\times 10^{-6} \, g^4_{BL} \left( \frac{M_P}{m_\sigma} \right) \, (x-x_{RH}). 
\end{eqnarray}
Then taking $Y(\infty)\simeq Y(x_{BL}=m_\sigma/M_{Z_{BL}})$, we estimate the DM relic density, 
\begin{eqnarray}
\Omega_{DM}h^2\simeq\frac{m_\sigma s_0 Y(\infty)}{\rho_0/h^2}\simeq 
   3.4 \times 10^{21} \, g^4_{BL} \, \left(\frac{m_\sigma}{{\rm 1 \, GeV}}\right) \left(\frac{{\rm 1 \, GeV}}{M_{Z_{BL}}}\right), 
\label{Omega}
\end{eqnarray}
where $s_0=2890/{\rm cm}^3$ is the entropy density of the present universe, 
  and $\rho_c/h^2=1.05 \times 10^{-5} \, {\rm GeV}{{\rm cm}^3}$ is the critical density.
This leads to the following expression for $g_{BL}$:
\begin{eqnarray}\label{gBL}
   g_{BL}\simeq 2.4 \times 10^{-6}\left(\frac{M_{Z_{BL}}}{{\rm 1 \, GeV}}\right)^{1/4}\left(\frac{{\rm 1 \, GeV}}{m_{\sigma}}\right)^{1/4}
\end{eqnarray}
to reproduce the observed DM relic density $\Omega_{DM}h^2 = 0.12$.

\begin{figure}[t]
  \centering
 \includegraphics[width=0.46\linewidth]{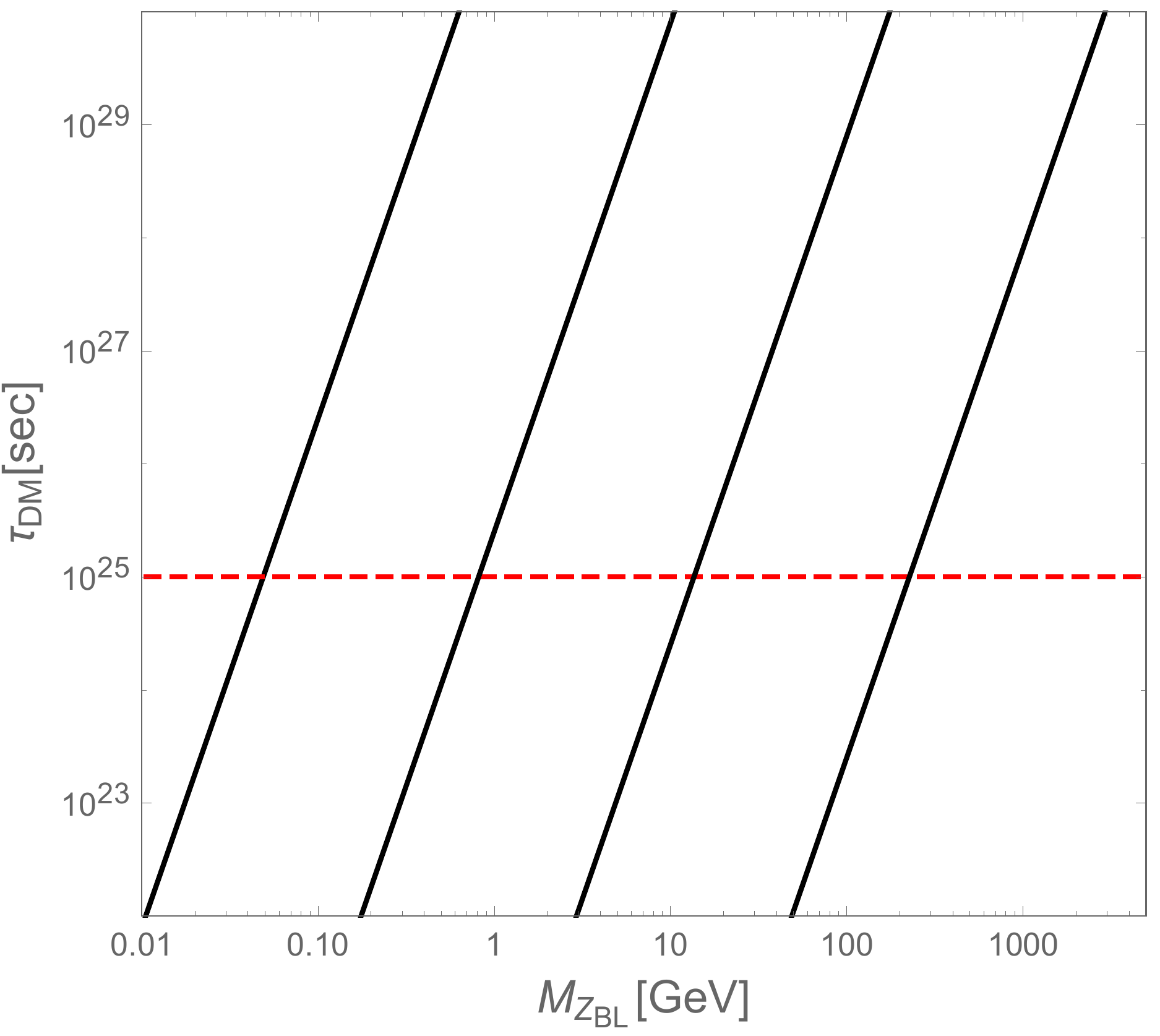}~~~
  \includegraphics[width=0.46\linewidth]{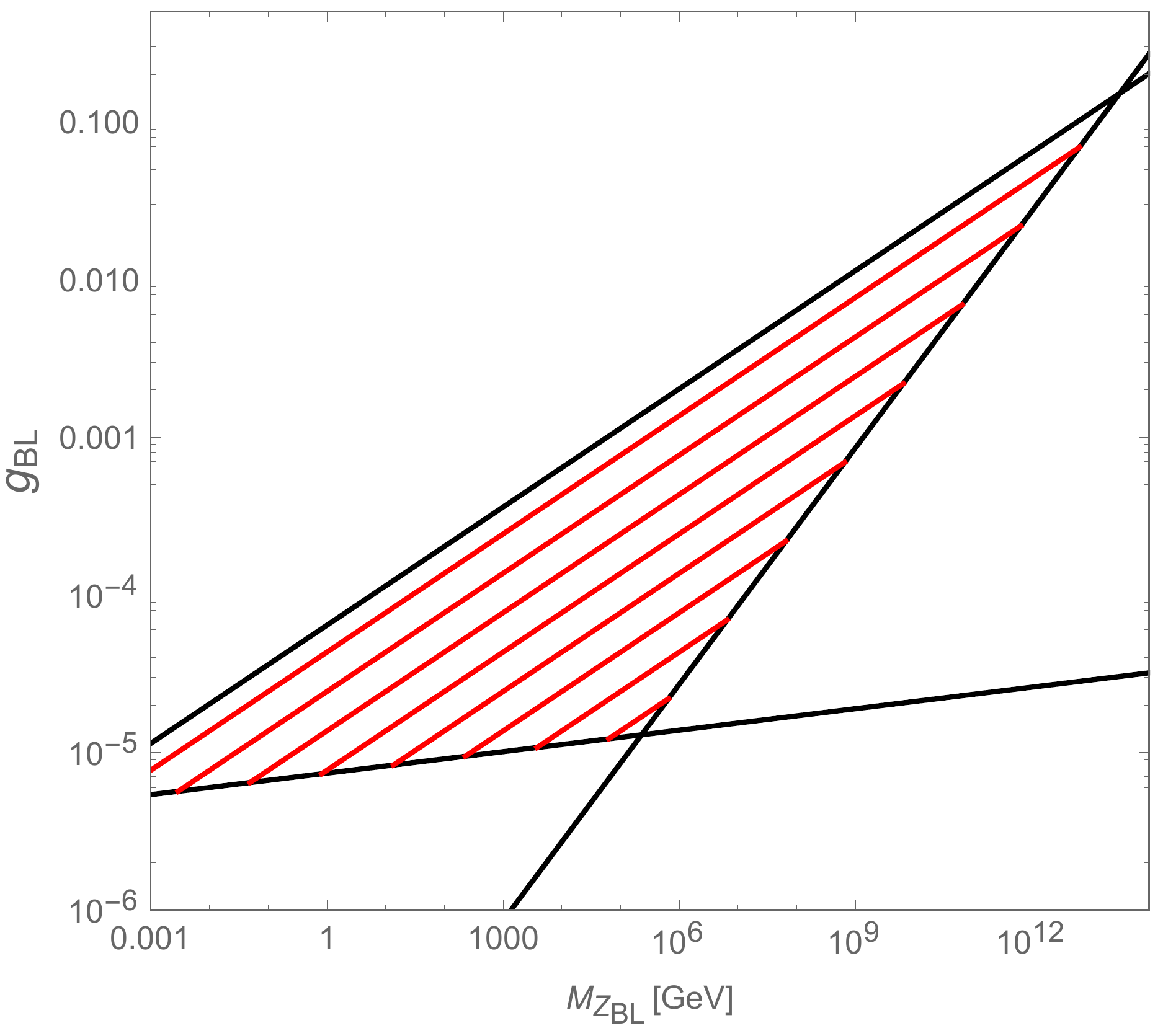}
  \caption{
{\bf Left Panel}: 
The dark matter $\sigma$ lifetime as a function of $M_{Z_{BL}}$. 
The diagonal solid lines correspond to  $m_\sigma$ =1 MeV, 10 MeV, 100 MeV, and 1 GeV  from left to right, 
along which the observed DM relic density of $\Omega_{DM} h^2= 0.12$ is reproduced. 
{\bf Right Panel}:
The $g_{BL}$ values as a function of $M_{Z_{BL}}$ from the requirement of relic density build-up. 
Different red lines correspond to different  DM masses 
($m_\sigma$ starting with 10 keV at the top and as we go below, we go in steps of a factor of 10 to 100 keV, 1 MeV, etc. till 100 GeV) 
that satisfy the relic density constraint i.e. $\Omega_{DM} h^2= 0.12$. 
Two diagonal black lines denote the condition of Eq.~(\ref{TH}), and the horizontal black line corresponds to Eq.~(\ref{Lmass}). 
}  
 \label{Fig:1}
\end{figure}

Using Eq.~(\ref{gBL}) in  Eq.~(\ref{LF}), we show the lifetime for various values of $m_\sigma$ in Fig.~\ref{Fig:1} (Left Panel).  
The diagonal lines from left to right correspond to $m_\sigma=1$ MeV, 10 MeV, 100 MeV and 1 GeV, respectively, 
  along which $\Omega_{DM}h^2 = 0.12$ is reproduced. 
The horizontal dashed line indicates the astrophysical bound on $\tau_\sigma > 10^{25}$ sec. 
Combining Eqs.~(\ref{LF2}) and (\ref{gBL}), we obtain a lower bound on $M_{Z_{BL}}$: 
\begin{eqnarray}
  M_{Z_{BL}} \gtrsim  210 \, \left(\frac{m_\sigma}{{\rm 1 \, GeV}}\right)^{11/9}~{\rm GeV}. 
\label{Lmass}  
\end{eqnarray}

\begin{figure}[t]
  \centering
 \includegraphics[width=0.6\linewidth]{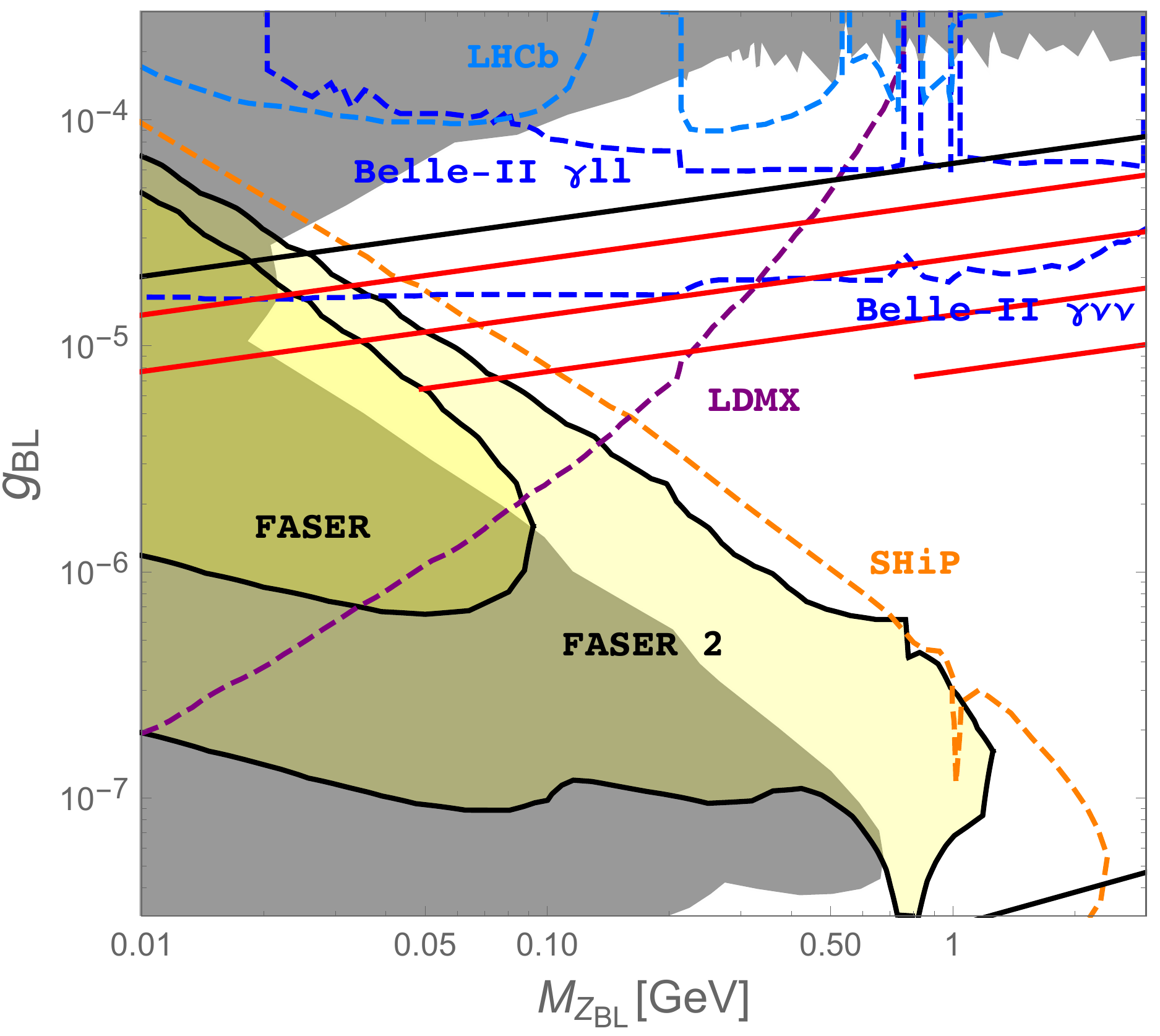}
  \caption{FASER reachable region of the parameter space of our model. 
The black lines at the top and bottom denote the upper and lower limits on the $g_{BL}$ (Eq.~(\ref{TH})). 
The red lines correspond to $m_\sigma=10$ keV, 100 keV, 1 MeV, and 10 MeV from top to bottom, respectively, 
along which $\Omega_{DM}=0.12$ is satisfied. 
The parameter region of 10 keV $\lesssim m_\sigma \lesssim 1$ MeV and 10 MeV$\lesssim M_{Z_{BL}} \lesssim$ a few GeV 
   can be tested by various Lifetime Frontier experiments in the near future. 
}
 \label{Fig:2}
\end{figure}

Considering all the constraints from Eqs.~(\ref{TH}),  (\ref{gBL}) and (\ref{Lmass}),
   we show the allowed parameter region in Fig.~\ref{Fig:1} (Right Panel).  
The region between two diagonal black lines satisfies the condition of Eq.~(\ref{TH}),  
   and the horizontal black line corresponds to Eq.~(\ref{Lmass}). 
The observed $\Omega_{DM}h^2 = 0.12$ is reproduced along the red lines
  each of which corresponds to a fixed $m_\sigma$ value. 
In the right panel, the region for $M_{Z_{BL}} \lesssim 10$ MeV and $g_{BL} \sim 10^{-5}$
  is excluded by the long-lived $Z_{BL}$ boson search results. 
See Fig.~\ref{Fig:2} for details.

In the right panel of Fig.~\ref{Fig:1}, we can see that there is an allowed parameter region 
  for $g_{BL} ={\cal O}(10^{-5})$ and $M_{Z_{BL}}=1$ MeV$-1$ GeV. 
For the parameter region, $Z_{BL}$ boson can be long-lived and such a long-lived neutral particle 
 can be explored in the near future by the Lifetime Frontier experiments, such as 
 FASER \cite{faser}, SHiP \cite{SHiP},  LDMX \cite{LDMX}, Belle II \cite{B2}, and LHCb \cite{LHCb1, LHCb2}.  
The $Z_{BL}$ boson search of the FASER experiment at the LHC is summarized in Ref.~\cite{faser} 
   along with the search reaches of other experiments as well as the current excluded region \cite{Bauer:2018onh}. 
In Fig.~\ref{Fig:2}, we show our results of the right panel of Fig.~\ref{Fig:2} 
   along with the summary plot in Ref.~\cite{faser}. 
The red lines correspond to $m_\sigma=10$ keV, 100 keV, 1 MeV, and 10 MeV from top to bottom, respectively. 
The parameter region of 10 keV $\lesssim m_\sigma \lesssim 1$ MeV and 10 MeV $\lesssim M_{Z_{BL}} \lesssim$ a few GeV 
   can be tested by various Lifetime Frontier experiments in the near future.

Before moving on to the next section, we comment on the dark matter production processes involving the RHN. 
If the RHN is in thermal equilibrium, the DM particles can also be created through $NN \to \sigma \sigma$. 
The estimate of $Y(\infty)$ from this process is analogous to the process $Z_{BL} Z_{BL} \to \sigma \sigma$, 
  and resultant density is roughly given by Eq.~(\ref{Omega}) with replacing $g_{BL} \to f$ and $M_{Z_{BL}} \to M_N$. 
Thus, we take $M_N < M_{Z_{BL}}$, or equivalently $f < g_{BL}$, so that the RHN mediated DM production becomes subdominant. 
Calculations for other processes such as $NN \to Z_{BL} \sigma$ and  $N Z_{BL} \to N \sigma$ are also analogous, 
and we can arrive at the same conclusion. 
We can also consider DM production processes through Dirac Yukawa couplings ($h_{SM}$) such as 
  $N  \, H \to  \ell  \, \sigma$ and $H \, \ell \to N \, \sigma$, 
  where $H$ and $\ell$ are the Higgs and lepton doublets, respectively. 
The DM productions can be subdominant if $h_{SM}$ is sufficiently small, in other words, 
through the seesaw formula, $N$ is sufficiently light. 
The discussion for the DM production process of $H \, \ell \to N \, \sigma$ is applicable
even if the RHN is not in thermal equilibrium.

\section{5.  PeV dark matter from $B-L$ breaking} 
So far we have explored the lower mass range of the dark matter. 
In this section, we explore the possibility that the $\sigma$ mass is in the PeV range so that one could attempt
   to explain the 100 TeV to PeV neutrinos observed in IceCube Neutrino Observatory \cite{Aartsen:2013bka} 
   by using $\sigma$ decay. 
We do not attempt to explain the IceCube signal here but simply to raise the possibility that a PeV mass $\sigma$ 
   can also qualify as the dark matter in our model in a different parameter range.
For this purpose, let us go through all the constraints on the model discussed above for this case.

\subsection{5.1 Lifetime constraint} 
This constraint is same as in the case of light $\sigma$ in Eq.~(\ref{lifetime}) except that in the right-hand side, 
  the masses of $\sigma$ and $Z_{BL}$ are now higher  and the new constraint can be written as
\begin{eqnarray}\label{lifetime1}
g_{BL} \leq 4.2 \times 10^{-8} \, \left(\frac{M_{Z_{BL}}}{{\rm 1 \, PeV}}\right) \left(\frac{{\rm 1 \, PeV}}{m_\sigma}\right)^{7/6}
\end{eqnarray}
If we restrict the $B-L$ breaking VEV $v_{BL}\leq 10^{16}$ GeV, then the lifetime  constraint can be translated to $M_{Z_{BL}}\sim 10^{10}$ GeV for $g_{BL}$ as large as $10^{-5}$.

We note that the one-loop $\sigma-h$ mixing contribution in this case leads to a very strong upper limit on the $g_{BL}$ value and much too small to generate enough relic density for the dark matter. In this case tyherefore, we fine tune the tree-level and one-loop $\sigma$-Higgs coupling to zero.
\subsection{5.2  Relic density constraints} 
We next explore the constraints of relic density on the heavy DM case.  
For such low $g_{BL}$ values, a heavy PeV scale DM and the $10^{10}$ GeV or higher mass $Z_{BL}$ would never have been in equilibrium.  The relic density must arise as in the first case via the freeze-in mechanism. 
Since $Z_{BL}$ is not in thermal equilibrium, the production takes place via the process $f\bar{f} \to Z_{BL} \sigma$ 
  through the SM fermion pair annihilations in the thermal plasma. 
In this case, the Boltzmann equation is given by 
\begin{eqnarray}
  \frac{dY}{dx} \simeq \frac{ \langle \sigma v \rangle}{x^2}\frac{s(m_\sigma)}{H(m_\sigma)} Y_{eq}  Y_{eq}^{BL}, 
\end{eqnarray}
where $Y_{eq}^{BL}$ is the yield of $Z_{BL}$ in thermal equilibrium and the cross section
  for the process $f\bar{f}\to Z_{BL}\sigma$ is estimated as 
\begin{eqnarray}
   \langle \sigma v \rangle = \frac{g^4_{BL}}{4\pi}\frac{M^2_{Z_{BL}}}{m^4_\sigma} x^4. 
\end{eqnarray}
Recall that the DM production stops at $T\simeq M_{Z_{BL}}$ due to kinematics. 
Using $Y_{eq}^{BL} \simeq 2 Y_{eq}$ for $T \gtrsim M^2_{Z_{BL}} \gg m_\sigma$, 
   we integrate the Boltzmann equation from $x_{RH}$ to $x_{BL}=\frac{m_\sigma}{M_{Z_{BL}}}$ and obtain
\begin{eqnarray}
   Y(x_{BL}) &\simeq & 3.4 \times 10^{-6} \, g^4_{BL} 
\left(\frac{M_{Z_{BL}}}{m_\sigma} \right)^2
\left( \frac{M_P}{m_\sigma} \right) \, (x_{BL}^3-x_{RH}^3) \nonumber \\
& \simeq &
3.4 \times 10^{-6} \, g^4_{BL} \left( \frac{M_P}{M_{Z_{BL}}} \right), 
\end{eqnarray}
where we have used $Y(x_{RH})=0$ and $x_{RH} \gg x_{BL}$. 
We now use, as before, $Y(\infty)\simeq Y(x_{BL})$ and estimate the DM relic density, 
\begin{eqnarray}
\Omega_{DM}h^2\simeq\frac{m_\sigma s_0 Y(\infty)}{\rho_0/h^2}\simeq 
   2.3 \times 10^{21} \, g^4_{BL} \, \left(\frac{m_\sigma}{{\rm 1 \, GeV}}\right) \left(\frac{{\rm 1 \, GeV}}{M_{Z_{BL}}}\right). 
\end{eqnarray}
In order to reproduce $\Omega_{DM} h^2=0.12$, we find 
\begin{eqnarray}
   g_{BL}\simeq 2.7\times 10^{-6}\left(\frac{M_{Z_{BL}}}{m_\sigma}\right)^{1/4}. 
\label{PeVDM}   
\end{eqnarray}
We require that the $Z_{BL}$ is not in equilibrium which gives the consistency condition 
\begin{eqnarray}
   g_{BL} <  2.7\times 10^{-8}\left(\frac{M_{Z_{BL}}}{{\rm GeV}}\right)^{1/2}. 
\label{out_eq}  
\end{eqnarray}

\begin{figure}[t]
  \centering
 \includegraphics[width=0.46\linewidth]{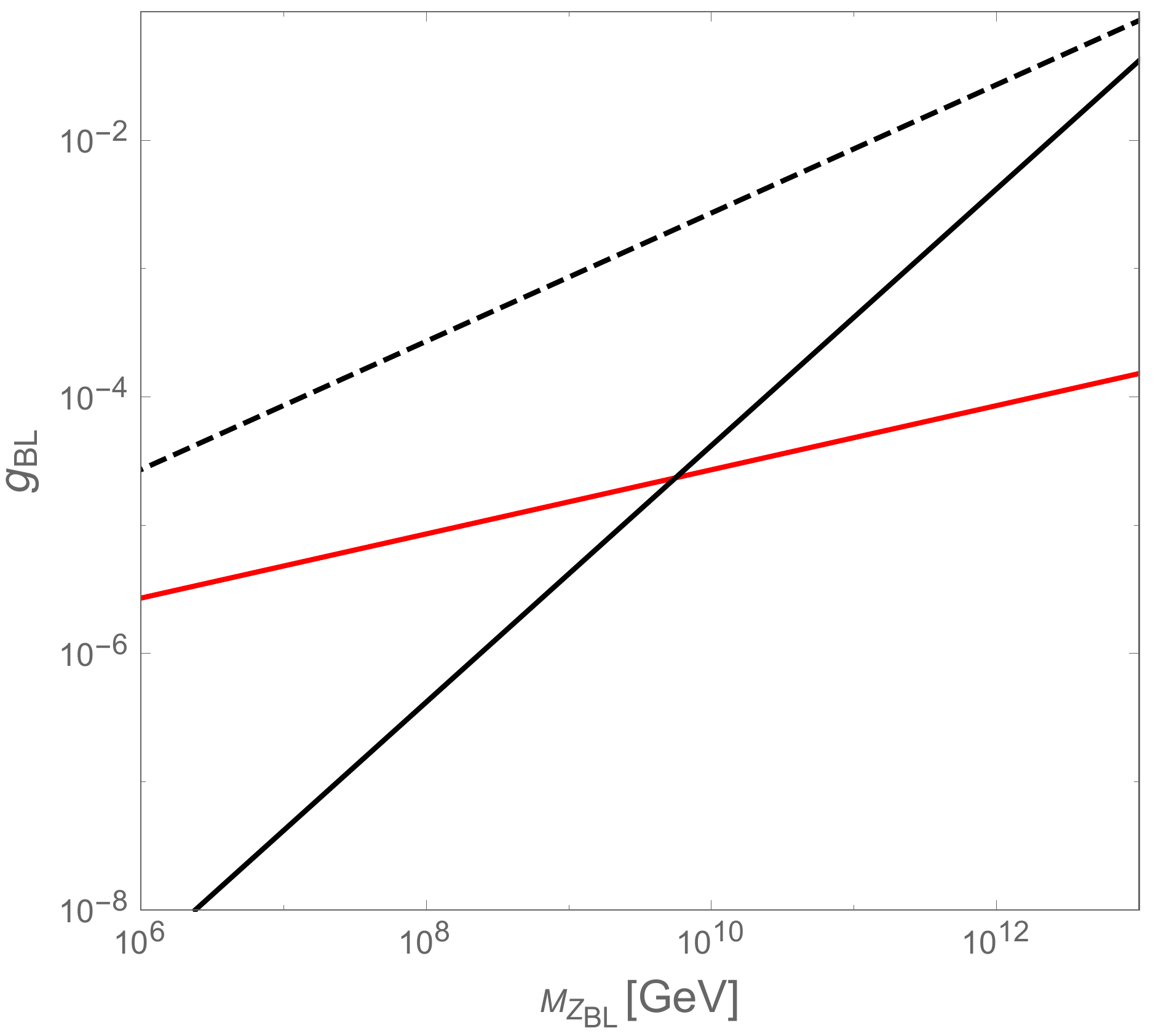}~~~
 \includegraphics[width=0.46\linewidth]{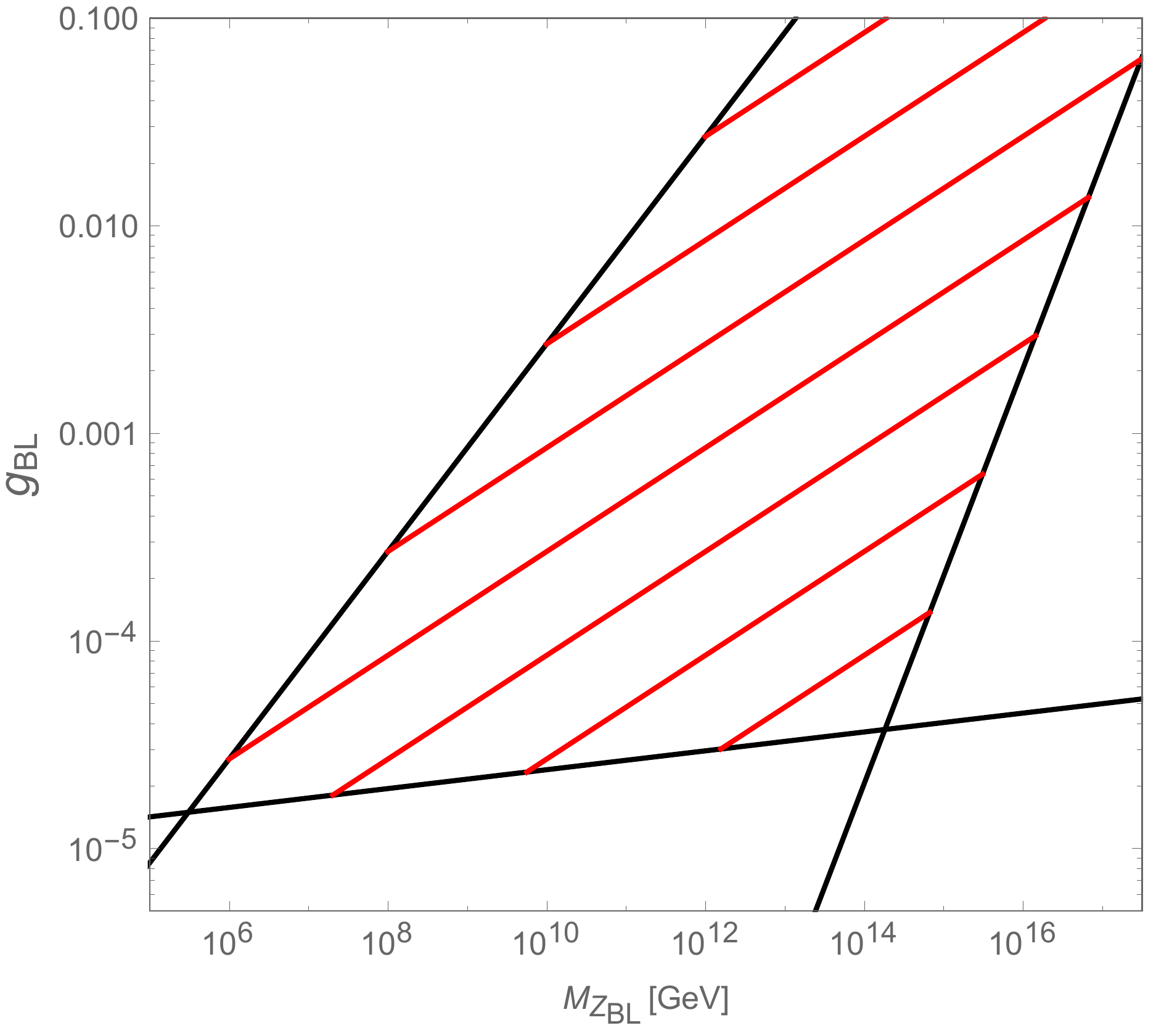} 
\caption{
{\bf Left Panel}: The red line corresponds to DM mass $m_\sigma= 1$ PeV with $\Omega_{DM} h^2 = 0.12$. 
This corresponds to the case where the DM is produced by $f \, \bar{f}\to Z_{BL} \, \sigma$. 
The left of dashed line corresponds to the DM being in thermal equilibrium and therefore is not the area for freeze-in case. 
The left of black solid line corresponds to $\tau_\sigma < 10^{25}$ sec. and is excluded. 
{\bf Right Panel}:
The red lines represent the DM masses from top 100 keV, 10 MeV, 1 GeV (jump of 100 times) till 100 PeV being the lowest red line.  
Along the red line $\Omega_{DM} h^2 =0.12$ is satisfied. The lower black line comes from the DM lifetime lower limit. Upper black line corresponds to $Z_{BL}$ not being in equilibrium.
The condition of  $v_{BL} \leq M_P$ is depicted by the right diagonal black line. 
}
\label{Fig:3}
\end{figure}

In Fig.~\ref{Fig:3} (Left Panel), we show our result for $m_\sigma=1$ PeV. 
The dashed line denotes the upper bound on $g_{BL}$ from the out-of-equilibrium condition of Eq.~(\ref{out_eq}). 
The diagonal black line shows the lifetime constraint of Eq.~(\ref{LF2}). 
Along the red line, the observed DM relic density is reproduced (see Eq.~(\ref{PeVDM})). 
In the figure, we find the lower bound on $M_{Z_{BL}}=4.5 \times 10^9$ GeV.    
In the right panel of Fig.~\ref{Fig:3}, we show the results for various values of $m_\sigma$. 
The red lines from top to bottom correspond to the results 
  for $m_\sigma=100$ keV, 10 MeV, 1 GeV, 100 GeV, 10 TeV, 1 PeV, and 100 PeV, respectively. 
The left diagonal black line denotes the out-of-equilibrium condition of Eq.~(\ref{out_eq}), 
  while the horizontal line depicts the lower bound on $M_{Z_{BL}}$ from the lifetime constraint 
  for various fixed values of $m_\sigma$. 
We also impose a condition of  $v_{BL} \leq M_P$, which is depicted by the right diagonal black line. 
We thus see that there is enough parameter range in the model for the dark matter to be in the PeV range
   so that it can be relevant to the PeV neutrinos observed in IceCube experiment. 
This is possible for $M_{Z_{BL}} \gtrsim 10^{10}$ GeV and $v_{BL}\gtrsim 10^{14}$ GeV.

\section{6. Prospects for SO(10) embedding} 
In this section, we like to point out that a slight variation of the model leads to its possible embedding into $SO(10)$ grand unified theory (GUT), which we believe should add to its theoretical appeal as a minimal GUT model that unifies neutrino masses and dark matter. The starting point of this discussion is the observation  that the hypercharge generator $Y$ is a linear combination of the $I_{3R}$ and the normalized $B-L$ generators $I_{BL}$ of $SO(10)$ as follows:
\begin{eqnarray}
Y=I_{3R}+\sqrt{\frac{2}{3}}I_{BL}
\end{eqnarray}
where $I_{BL}=\sqrt{\frac{3}{2}} \frac{B-L}{2}$. The $B-L$ generator in the main body of the paper is not orthogonal to the $Y$ generator defined above. Therefore, it cannot emerge from $SO(10)$ breaking since $I_{BL}$ is not orthogonal to $Y$ defined above.  
Instead if we consider the generator $\tilde{I}\equiv -4I_{3R}+3(B-L)$,  we get ${\rm Tr}(\tilde{I}Y)=0$  (i.e. they are orthogonal) for any irreducible representation of $SO(10)$ and can therefore emerge from $SO(10)$ breaking. 
This generator was also identified in Ref~\cite{Nobu1} as the generator $U(1)_X$ for $x_H= -4/5$. Indeed, it has been shown in Ref.~\cite{diluzio} that such a generator emerges out of $SO(10)$ breaking by a {\bf 45} Higgs field. To see this note that {\bf 45} Higgs under $SU(3)\times SU(2)_L\times SU(2)_R\times U(1)_{B-L}$ group has multiplets $(1,1,1,0)$ and $(1,1,3,0)$ which can take 
VEVs $\omega_Y$ and $\omega_{BL}$, respectively. 
If we fine-tune the parameters of the Higgs  potential, we can get $\omega_Y=\omega_{BL}$
in which case the unbroken generators are $U(1)_Y\times U(1)_{\tilde{I}}$. 
The normalized $\tilde{I}=\frac{1}{2\sqrt{10}}(-4I_{3R}+3(B-L))$. 

 As it turns out, the dark matter phenomenology discussed above remains unchanged if we use the Higgs field $\sigma$ to break the $U(1)_{\tilde{I}}$ symmetry. The $\sigma$ field then emerges from the {\bf 126}-dimensional representation of $SO(10)$ and our dark matter field $\sigma$ has $\tilde{I}=\frac{\sqrt{10}}{2}$ and therefore has all the  properties required above for our dark matter.

 Our scenario for $SO(10)$ breaking is as follows: we use {\bf 45}-dimensional Higgs field to break $SO(10)$ down to $SU(5)\times U(1)_{\tilde{I}}$ by choosing the vacuum with $\omega_Y=\omega_{BL}$, as noted above. The $\tilde{I}$ quantum numbers of fermions are then given by $\tilde{I}({\bf 10})= \frac{1}{2\sqrt{10}}$, $\tilde{I}({\bar{\bf 5}})= \frac{-3}{2\sqrt{10}}$ and $\tilde{I}({\bf 1})= \frac{5}{2\sqrt{10}}$, where {\bf 10}, ${\bar{\bf 5}}$ and {\bf 1} are the $SU(5)$ representations in $SO(10)$ spinor {\bf 16}. 
For a {\bf 10}-representation Higgs field in $SO(10)$, which is decomposed into ${\bf 5}+{\bar{\bf5}}$ under $SU(5)$ 
and includes the SM Higgs doublet,  the $\tilde{I}$ quantum numbers are given by 
$\tilde{I}({\bf 5})= \frac{-2}{2\sqrt{10}}$ and $\tilde{I}({\bf 5})= \frac{2}{2\sqrt{10}}$. 

Let us now discuss the evolution of the $\tilde{I}$ gauge coupling. 
The evolution of $U(1)_{\tilde{I}}$ gauge coupling ($g_{\tilde I}$) is given by
 \begin{eqnarray}
 \mu \frac{d \alpha^{-1}_{\tilde I}}{d \mu}=\frac{b_{\tilde I}}{2\pi}, 
 \end{eqnarray}
where $b_{\tilde I}=-49/10$ at a scale $\mu$ below the $SU(5)$ unification while $b_{\tilde I}=-5$
in $SU(5) \times U(1)_{\tilde{I}}$ theory by considering that the SM Higgs doublet is embedded
into a {\bf 5}-representation in $SU(5)$.  
For simplicity, we have assumed that in each step of the gauge symmetry breaking, 
$SO(10) \to SU(5) \times U(1)_{\tilde{I}} \to SU(3)_c \times SU(2)_L\times U(1)_Y \times U(1)_{\tilde{I}}
\to SU(3)_c \times SU(2)_L\times U(1)_Y$, only the minimal sets of Higgs fields are light. 

To see our coupling unification strategy in this model,  we first discuss the $SU(5)$ unification without supersymmetry. 
As is clear, in this case, we will need extra fields beyond the SM fields below the $SU(5)$ unification scale. 
For this purpose, we introduce $n_3$ real scalar $SU(2)_L$ triplets with $Y=0$ and $n_8$ real scalar color octets with $Y=0$. 
The coupling evolution equations in this case are the following: 
\begin{eqnarray}
\mu \frac{d\alpha^{-1}_1}{d \mu}&=&-\frac{1}{2\pi} \left( \frac{41}{10} \right),   \nonumber\\
\mu \frac{d\alpha^{-1}_2}{d \mu}&=& \frac{1}{2\pi} \left( \frac{19}{6} - \frac{n_3}{3}\theta(\mu-M_3) \right),   \nonumber\\
\mu \frac{d\alpha^{-1}_3}{d \mu}&=& \frac{1}{2\pi} \left( 7 - \frac{n_8}{2}\theta(\mu-M_8) \right), 
\end{eqnarray}
where $M_{3,8}$ stand for the masses of the triplet $({\bf 1}, {\bf 3},1)$ and octet  $({\bf 8}, {\bf 1}, 0)$ fields, respectively. 
Solving these equations with $n_3=5$ with mass $M_3=5$ TeV and $n_8=3$ with mass $M_8=200$ TeV, 
we find that the $SU(5)$ gauge coupling unification is achieved at $M_U=6.8 \times 10^{15}$ GeV.

\begin{figure}[t]
  \centering
 \includegraphics[width=0.7\linewidth]{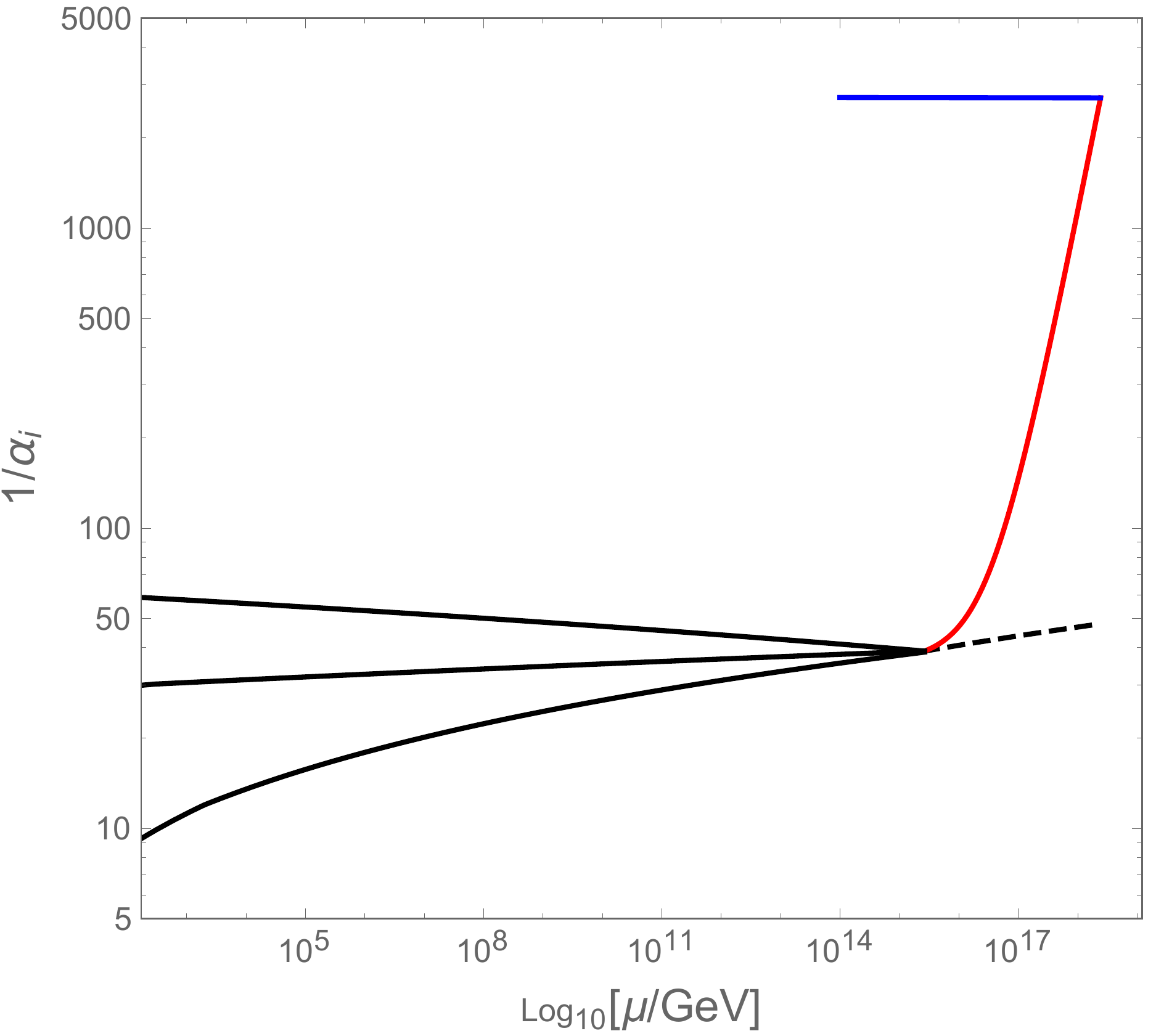}
  \caption{
Unification of gauge couplings in the presence of one extra dimension. 
The horizontal blue line denotes $\alpha_{\tilde I}^{-1}$ while solid black lines
 from top to bottom denote $\alpha_1^{-1}$, $\alpha_2^{-1}$ and $\alpha_3^{-1}$, respectively. 
Here, we  have set the $U(1)_{\tilde I}$ gauge boson mass (corresponding to $M_{Z_{BL}}$ in the previous sections) 
  to be $10^{14}$ GeV as an example. 
The red curve represents the running of $\alpha_5^{-1}$ in the presence of the gauge boson KK modes.    
For a comparison with 4-dimensional theory, we show the dashed line for the $SU(5)$ without the KK mode contributions. 
 }
 \label{Fig:GCU}
\end{figure}

Let us now proceed to $SO(10)$ unification i.e. the running of the $g_{\tilde I}$ coupling from its breaking scale (which does not affect very much) to where it unifies with the $SU(5)$ coupling evolving after the $SU(5)$ unification scale. 
We see that due to the small value of $g_{\tilde I}$ required to get the relic density from the freeze-in mechanism, 
the $SO(10)$ gauge coupling unification in 4-dimensions is hard to obtain. 
We therefore assume that above the $SU(5)$ GUT scale, the model becomes five dimensional \cite{keith} 
with the fifth dimension compactified on $S^1/Z_2$ orbifold with a radius $R={M_U}^{-1}$. 
In that case if we assume that the gauge fields are in the bulk while all the matter and Higgs fields are on a brane at an orbifold fixed point, 
their Kaluza-Klein (KK) modes contribute to the running of the $SU(5)$ coupling whereas $U(1)_{\tilde{I}}$ being abelian its coupling running does not get any extra contribution from the opening of fifth dimension. 
The evolution of the $SU(5)$ gauge coupling ($\alpha_5$) obeys 
\begin{eqnarray}
\mu \frac{d\alpha^{-1}_5}{d \mu}= \frac{1}{2\pi} 
\left( \frac{43}{3} - \frac{1}{6} -\frac{5}{6} \left(1+n_3+n_8 \right) 
+ \frac{55}{3} \sum_{n=1} \theta(\mu-\sqrt{1+n^2} M_U)
\right). 
\end{eqnarray}
Here, in the parenthesis of the right-hand side, 
$43/3$ is the contribution from the zero-mode $SU(5)$ gauge boson and the SM fermions, 
$-1/6$ from the ${\bf 5}$-representation Higgs field, and 
$-\frac{5}{6} \left(1+n_3+n_8 \right)$ from one adjoint Higgs to break the $SU(5)$ symmetry
and $n_3+n_8$ adjoint Higgs field into which the triplet and octet scalars are  embedded, 
and the last term is the contribution from the $SU(5)$ gauge boson KK modes. 
For the KK mode mass spectrum, we have simply added the contribution from the $SU(5)$ symmetry breaking. 
Once the extra dimension opens, the contribution from the KK modes changes the scale dependence
of the running gauge coupling from a log to a power \cite{keith}. 
Thus it is possible to unify the $SU(5)$ and $U(1)_{\tilde{I}}$ couplings into $SO(10)$ coupling as desired. 
This is shown in Fig.~\ref{Fig:GCU}.
In the figure, the $SO(10)$ gauge coupling unification is achieved at $M_P$ with a unified coupling $g_{SO(10)} \simeq 0.1$. 
This result corresponds to an allowed parameter set, $m_\sigma \simeq 100$ keV and $M_{Z_{BL}}=10^{14}$ GeV, 
in the right panel of Fig.~\ref{Fig:3}.

As far as proton decay is concerned, the primary mode is $p\to e^++\pi^0$ mediated by the $SU(5)$ gauge boson. 
The proton decay amplitude gets contribution from all the KK excitations of the SU(5) gauge fields, 
and we estimate the modification of a coefficient of the 4-Fermi operator to be 
\begin{eqnarray}
   \frac{1}{M_U^2} \to  \frac{1}{M_U^2}  \left(1 + \sum_{n=1}^\infty \frac{1}{1+n^2}    \right)
   \simeq \frac{2.08}{M_U^2} \equiv \frac{1}{\Lambda^2}. 
\end{eqnarray}
Then, (ignoring threshold effects) the proton lifetime is estimated as
\begin{eqnarray}
   \tau_p \simeq \frac{\Lambda^4}{\alpha_U^2 m_p^5}, 
\end{eqnarray}
where $m_p=0.938$ GeV. 
Using $\alpha_5(M_U) \simeq 0.026$ and $M_U \simeq 6.8 \times 10^{15}$ GeV from Fig.~\ref{Fig:GCU}, 
we find that $\tau_p\simeq 2.1 \times 10^{34}$ years, 
which is consistent with the lower bound $\tau_p \geq1.6 \times 10^{34}$ years 
from the Super-Kamionkande results \cite{Miura:2016krn}. 
More importantly, we would expect that $p\to e^+\pi^0$ should be observable in the next round of proton decay searches
at Hyper-Kamiopkande \cite{Abe:2011ts} or the model will be ruled out.

\section{7. Concluding remarks} 
We have presented a minimal model based on a $U(1)_{B-L}$ extension of the standard model where the $B-L$ breaking Higgs field plays the role of a decaying dark matter. We discuss two regions of the DM masses: one light mass region in the keV to MeV range and another where the DM mass is in the PeV range. 
In both cases, due to the stability requirement of the Dark matter, the freeze-in mechanism is required to understand the observed relic density of DM. We then discuss how the model can be tested in the FASER and other Lifetime Frontier experiments. Finally, we show how the model can emerge from an $SO(10)$ GUT model. Coupling unification in this case requires that the model be part of a five dimensional space-time with the compactification radius being of the order of the inverse of the $SU(5)$ unification scale $M_U$. 
This embedding reflects itself in an enhanced decay rate for the proton due to extra gauge KK mode contributions, which we have estimated. The model may have TeV scale hypercharge neutral weak iso-triplet and color octet scalars, which have interesting LHC phenomenology \cite{triplet,octet}. Discussion of this phenomenology is beyond the scope of this paper. There are also ranges for the RHN masses in the model where resonant leptogenesis can generate the baryon asymmetry of the universe. This will be the subject of a forthcoming publication.

\section*{Acknowledgement} 
The work of R.N.M. is supported by the National Science Foundation grant No.~PHY-1620074 and PHY-1914631,  
and the work of N.O. is supported by the US Department of Energy grant No.~DE-SC0012447.

\end{document}